\documentclass[conference]{IEEEtran}

\usepackage{graphicx}
\usepackage{amsmath,bbm,epsfig,amssymb,amsfonts,  amstext, verbatim,amsopn,cite,subfigure,multirow,multicol,lipsum}
\usepackage{balance}
\usepackage{url}
\usepackage{amsfonts}
\usepackage{epsfig}
\usepackage{setspace}
\usepackage{stmaryrd}
\usepackage{psfrag}	
\usepackage{multirow}
\usepackage{float}
\usepackage[process=auto]{pstool}
\usepackage{etoolbox}
\usepackage{algorithm}
\usepackage{algorithmic}
\usepackage{hyperref}

\usepackage{pifont}
\allowdisplaybreaks

\newcommand*{\Scale}[2][4]{\scalebox{#1}{$#2$}}%
\newcommand*{\QED}{\hfill\ensuremath{\square}}%

\newtheorem{theo}{Theorem}

\newtheorem{remk}{Remark}
\newtheorem{defin}{Definition}
\newtheorem{prop}{Proposition}
\newtheorem{corol}{Corollary}
\newtheorem{examp}{Example}

\newtoggle{ConfVersion}

\togglefalse{ConfVersion}

\iftoggle{ConfVersion}{%
  
}{%
  
}

\EndPreamble
\begin{document}

\title{SCW Codes for Optimal CSI-Free Detection in Diffusive Molecular Communications \vspace{-0.3cm}}

\author{Vahid Jamali$^\dag$, Arman Ahmadzadeh$^\dag$, Nariman Farsad$^\ddag$, and Robert Schober$^\dag$\\
\IEEEauthorblockA{\dag Friedrich-Alexander University (FAU), Erlangen, Germany,~~
\ddag Stanford University, Stanford, California, USA \vspace{-0.5cm}
}
}

\maketitle

\begin{abstract}
Instantaneous or statistical channel state information (CSI) is needed for most detection schemes developed  in the molecular communication (MC) literature. Since the MC channel changes, e.g., due to variations in the velocity of  flow, the temperature, or the distance between transmitter and receiver,  CSI acquisition has to be conducted repeatedly to keep track of CSI variations. Frequent CSI acquisition  may entail a large overhead whereas infrequent CSI acquisition may result in a low CSI estimation quality. To cope with these issues,  we design codes  which facilitate maximum likelihood sequence detection at the receiver without \textit{instantaneous} or \textit{statistical} CSI. In particular, assuming concentration shift keying modulation, we show that a class of codes, referred to as \emph{strongly constant-weight (SCW) codes}, enables \textit{optimal CSI-free} sequence detection at the cost of decreasing the data rate. For the proposed SCW codes, we analyze the code rate and the error rate. Simulation results verify our analytical derivations and reveal that the proposed CSI-free detector for SCW codes outperforms the baseline coherent and non-coherent detectors for uncoded transmission\iftoggle{ConfVersion}{}{\footnote{This is an extended version of a paper available in Proc. IEEE ISIT 2017.}}.

\end{abstract}

\section{Introduction}

In contrast to conventional wireless communication systems that encode data into electromagnetic waves, synthetic molecular communication (MC) systems are envisioned to embed data into the characteristics of  signaling molecules such as their concentration, type, and time of the release  \cite{Nariman_Survey,Survey_Mol_Net}. Diffusive MC  is a common  strategy for communication between nano-/microscale entities in nature such as bacteria, cells, and organelles (i.e., components of cells) \cite{BioPhysic}. Therefore, diffusive MC has been considered as a bio-inspired approach for communication between small-scale nodes where conventional wireless communication may be inefficient or even infeasible \cite{Nariman_Survey,Survey_Mol_Nono}.

In diffusive MC, the expected number of signalling molecules observed at the receiver at a given time after the emission of a known number of molecules by the transmitter and the expected number of interfering molecules observed at the receiver constitute the channel state information (CSI) \cite{NanoCOM16}. Knowledge of the \textit{instantaneous} CSI is needed in general for optimal coherent detection \cite{HamidJSAC} and can be obtained using  training sequence-based channel estimators \cite{TCOM_MC_CSI}. The CSI of an MC channel depends on  various parameters such as the diffusion coefficient of the signaling molecules, the velocity of the flow in the channel, the concentration of enzyme degrading the signaling molecules, the distance between the transmitter and the receiver, etc., see \cite[Chapter~4]{BioPhysic}, \cite{ArmanMobileMC}. A change in any of these parameters affects the CSI of the considered MC channel. Therefore, CSI acquisition has to be  conducted repeatedly to keep track of CSI  variations. To reduce the CSI acquisition overhead, the authors in \cite{NanoCOM16} derived the optimal non-coherent detector which requires only \textit{statistical} CSI instead of instantaneous CSI. The statistical CSI for a particular MC channel can be estimated using empirical measurements. However, this may not always be  possible, particularly for practical MC systems with limited processing capabilities. In fact,  an experimentally verified statistical channel model for MC systems has not been reported yet. 

Motivated by the aforementioned challenges in CSI acquisition, in this paper, we propose a class of codes, referred to as \textit{strongly constant-weight (SCW) codes}, for which we show that maximum likelihood (ML) detection is possible without instantaneous or statistical CSI knowledge. In other words, SCW codes enable optimal CSI-free detection at the expense of a decrease in data rate. We analyze the code rate and the error rate of the proposed SCW codes. In addition, we study the properties of  the special cases of binary SCW codes and balanced~SCW~codes. 

We note that the problem considered in this paper, i.e., the design of SCW codes, can be seen as a modulation design, code design, or coded modulation design problem \cite{Ungerboeck_CodedMod,Huber_MultiLevCode}. However, our main motivation in employing SCW codes here is to devise an optimal ML detection algorithm that does not require CSI. We note that SCW codes  are special cases of the widely-known constant-weight (CW) codes \cite{WeightCodeClass,CW_qray}. However, to the best of our knowledge, SCW codes and their application for CSI-free detection have not been considered in the MC literature~yet.

\textit{Notations:} We use the following notations throughout this paper: $\mathsf{E}\{\cdot\}$ denotes expectation. Bold lower case letters  denote vectors and $\mathbf{a}^{\mathsf{T}}$ represents the transpose of vector $\mathbf{a}$. $H_n(\cdot)$ represents the entropy function for the logarithm to base $n$, $n!$ is the factorial of $n$, and $O(n)$ denotes the complexity order of $n$.  Moreover,  $\mathcal{P}(\lambda)$  denotes a Poisson random variable (RV) with mean  $\lambda$, $\lfloor\cdot\rfloor$ denotes the floor function which maps a real number to the largest integer number that is smaller or equal to the real number, and $\mathbf{1}\{\cdot\}$ is an indicator function that is equal to one if the argument is true, and is equal to zero otherwise.

\section{System Model}

We consider an MC system  consisting of a transmitter, a channel, and a receiver. We employ concentration shift keying (CSK) modulation  where the transmitter releases  $s[k]N^{\mathrm{tx}}$ molecules at the beginning of the $k$-th symbol interval to convey symbol $s[k]\in\mathcal{S}$ \cite{Nariman_Survey}. Here, $N^{\mathrm{tx}}$ is the maximum number of molecules that the transmitter can release in one symbol interval, i.e., a peak per-symbol ``power" constraint, and $\mathcal{S}=\{\eta_0,\eta_1,\dots,\eta_{L-1}\}$ denotes the symbol set  where $L$ is the number of available symbols. Without loss of generality, we assume $\eta_0<\eta_1<\cdots<\eta_{L-1}$, $\eta_0=0$, and $\eta_{L-1}=1$. Moreover, let $\mathbf{s}=[s[1],s[2],\dots,s[K]]^{\mathsf{T}}$  denote a codeword  comprising $K$ symbols. 


The released molecules diffuse through the fluid medium between the transmitter and the receiver. We assume that the movements of individual molecules are  independent from each other. The number of observed (counted) molecules at the receiver in each symbol interval is considered as the received signal. We assume perfect symbol synchronization \cite{ICC2017_MC_Arxiv}. Let $\mathbf{r}=[r[1],r[2],\dots,r[K]]^{\mathsf{T}}$ denote the vector of observations corresponding to sequence $\mathbf{s}$ where  $r[k]$ denotes the number of  molecules observed at the receiver  in symbol interval $k$. Due to the counting process at the receiver,  $r[k]$ can be accurately modelled as a Poisson RV, see \cite{NanoCOM16,Yilmaz_Poiss,HamidJSAC}, i.e.,
\begin{align} \label{Eq:ChannelInOut}
  r[k]  \sim \mathcal{P}(s[k]\bar{c}_{\mathrm{s}}+\bar{c}_{\mathrm{n}}), 
\end{align}
where   $\bar{c}_{\mathrm{s}}$ is the  number of  molecules \textit{expected} to be observed at the receiver  in symbol interval $k$ due to the release of $N^{\mathrm{tx}}$ molecules by the transmitter at the beginning of symbol interval~$k$ and  $\bar{c}_{\mathrm{n}}$ is the \textit{expected} number of interfering noise molecules comprising  multiuser interference (caused by other MC links) and  external noise (originating from natural sources) observed by the receiver \cite{NanoCOM16}. The inter-symbol interference (ISI) free communication model in (\ref{Eq:ChannelInOut}) implies that the symbol intervals are chosen large enough such that the channel impulse response (CIR) sufficiently decays to zero within one symbol interval.
We note that enzymes \cite{Adam_Enzyme} and reactive information molecules, such as acid/base molecules \cite{Acid_Base}, may be used to speed up the decaying of the CIR as a function of time, see \cite[Section~2]{NanoCOM16} for further justification. 

Note that the MC channel in (\ref{Eq:ChannelInOut}) is characterized by $\bar{c}_{\mathrm{s}}$ and $\bar{c}_{\mathrm{n}}$. Hence, we refer to the vector $\bar{\mathbf{c}}=[\bar{c}_{\mathrm{s}},\bar{c}_{\mathrm{n}}]^{\mathsf{T}}$ as the CSI of the considered MC system in the remainder of this paper. Moreover, we assume that the CSI remains unchanged over one block of transmitted symbols, i.e., one codeword, but may change from one block to the next (e.g., due to a change of the flow velocity or the distance between transmitter and receiver). To model this, we assume that the CSI, $\bar{\mathbf{c}}$, is an RV that takes its values in each block according to probability density function (PDF)  $f_{\bar{\mathbf{c}}}(\bar{c}_{\mathrm{s}},\bar{c}_{\mathrm{n}})$. For future reference, we define $\mathsf{SNR}=\frac{\bar{c}_{\mathrm{s}}}{\bar{c}_{\mathrm{n}}}$ as the signal-to-noise ratio (SNR).

\section{Optimal CSI-Free Detection Using SCW Codes}

In this section, we first introduce the class of SCW codes. Subsequently, we formulate the ML problems for coherent and non-coherent sequence detection which in general require instantaneous and statistical CSI, respectively. Finally, as the main result of this paper, we show that CSI-free ML detection is possible if SCW codes are adopted. For future reference, we define some auxiliary variables. In particular,  let $\omega(\mathbf{s})=\sum_{k=1}^K  s[k]$ denote the weight of sequence $\mathbf{s}$ and let $\omega_{\ell}(\mathbf{s},\mathbf{r})=\sum_{k=1}^K r[k] \mathbf{1}\{s[k]=\eta_{\ell}\}$ denote the weight of the observation sequence $\mathbf{r}$ corresponding to the positions where $s[k]=\eta_{\ell}$.
The definition of \textit{SCW codes} is formally presented in the following.
 
\begin{defin}\label{Def:StrConstCode}
SCW codes are denoted as $\boldsymbol{\mathcal{S}}^{\mathrm{sc}}(\bar{\boldsymbol{\omega}})$ with weight vector $\bar{\boldsymbol{\omega}}=[\bar{\omega}_0,\bar{\omega}_1,\dots, \bar{\omega}_{L-1}]^{\mathsf{T}}$, where for all codewords $\mathbf{s}$ in the codebook, the following property holds 
\begin{IEEEeqnarray}{lll} \label{Eq:StrWeightCode}
\sum_{k=1}^{K}\mathbf{1}\{s[k]=\eta_{\ell}\} = \bar{\omega}_{\ell},\quad \forall \eta_{\ell}\in\mathcal{S} \,\,\text{and}\,\,\forall \mathbf{s} \in \boldsymbol{\mathcal{S}}^{\mathrm{sc}}(\bar{\boldsymbol{\omega}}). 
\end{IEEEeqnarray}
An SCW code is called a full code if all possible codewords that satisfy  (\ref{Eq:StrWeightCode}) are included in the codebook. Moreover, an SCW code is called balanced if all weights $\bar{\omega}_{\ell}$ are identical, i.e., $\bar{\omega}_{\ell}=\bar{\omega},\,\,\forall \ell$ holds. 
\hfill \QED
\end{defin}

\begin{remk}
CW codes, denoted by $\boldsymbol{\mathcal{S}}^{\mathrm{c}}(K,\omega)$, have been widely employed in conventional communication systems \cite{WeightCodeClass,CW_qray}. For these codes, weight $\omega(\mathbf{s})=\omega$ is constant for all codewords in the codebook. Obviously, an SWC code $\boldsymbol{\mathcal{S}}^{\mathrm{sc}}(\bar{\boldsymbol{\omega}})$ is also a CW code $\boldsymbol{\mathcal{S}}^{\mathrm{c}}(K,\omega)$ with $K=\sum_{\ell=0}^{L-1}\bar{\omega}_{\ell}$ and $\omega=\sum_{\ell=0}^{L-1}\bar{\omega}_{\ell}\eta_{\ell}$. We note that for binary codes, i.e., $\mathcal{S}=\{0,1\}$, CW codes and SCW codes become equivalent, i.e., $\boldsymbol{\mathcal{S}}^{\mathrm{sc}}([\bar{\omega}_0,\bar{\omega}_1]^{\mathsf{T}})=\boldsymbol{\mathcal{S}}^{\mathrm{c}}(K,\omega)$ where $\omega=\bar{\omega}_1=K-\bar{\omega}_0$.
\end{remk}

\begin{remk}
In Section~II, we assumed that the maximum number of molecules that the transmitter can release in one symbol interval is limited to $N^{\mathrm{tx}}$, i.e., a peak power constraint. Hence, for CW/SCW codes, the number of  molecules released by the transmitter of the considered MC system is identical to $N^{\mathrm{tx}}\omega$ for all codewords. Therefore, for CW/SCW codes, the average number of  molecules released per symbol interval, denoted by $\bar{N}^{\mathrm{tx}}$,  is given by $\bar{N}^{\mathrm{tx}}=\frac{\omega}{K}N^{\mathrm{tx}}$.
\end{remk}

The  ML problems for coherent and non-coherent sequence detection are given by
\begin{align} 
  \hat{\mathbf{s}}^{\mathrm{c}}  &= 
  \underset{\mathbf{s}\in\boldsymbol{\mathcal{S}}}{\mathrm{argmax}} \,\, f_{\mathbf{r}}(\mathbf{r}|\bar{\mathbf{c}},\mathbf{s}),
   \label{Eq:ML}\\
   \hat{\mathbf{s}}^{\mathrm{nc}}  &= \underset{\mathbf{s}\in\boldsymbol{\mathcal{S}}}{\mathrm{argmax}} \,\, 
     \int_{\bar{c}_{\mathrm{s}}} \int_{\bar{c}_{\mathrm{n}}} f_{\mathbf{r}}(\mathbf{r}|\bar{\mathbf{c}},\mathbf{s}) f_{\bar{\mathbf{c}}}(\bar{c}_{\mathrm{s}},\bar{c}_{\mathrm{n}}) \mathrm{d}\bar{c}_{\mathrm{s}} \mathrm{d}\bar{c}_{\mathrm{n}}, \label{Eq:ML_NonCoherent}
\end{align}
respectively, where $\boldsymbol{\mathcal{S}}$ is the set of available sequences $\mathbf{s}$ and $f_{\mathbf{r}}(\mathbf{r}|\bar{\mathbf{c}},\mathbf{s})$ is the likelihood function conditioned on a given CSI vector, $\bar{\mathbf{c}}$, and a given hypothesis sequence $\mathbf{s}$. Exploiting the fact that the observations in different symbol intervals are independent, we obtain $f_{\mathbf{r}}(\mathbf{r}|\bar{\mathbf{c}},\mathbf{s})$ as 
\begin{align} 
f_{\mathbf{r}}(\mathbf{r}|\bar{\mathbf{c}},\mathbf{s}) = \prod_{k=1}^{K} \frac{\left( \bar{c}_{\mathrm{s}} s[k] + \bar{c}_{\mathrm{n}} \right)^{r[k]} \mathsf{exp}\left(- \bar{c}_{\mathrm{s}} s[k] - \bar{c}_{\mathrm{n}} \right)}{r[k]!}. \label{Eq:Likelihood}
\end{align}

In order to find the ML sequence for general $\boldsymbol{\mathcal{S}}$, the coherent sequence detection in (\ref{Eq:ML})   requires the instantaneous CSI, i.e., $(\bar{c}_{\mathrm{s}},\bar{c}_{\mathrm{n}})$, whereas the non-coherent sequence detection in (\ref{Eq:ML_NonCoherent})  requires  the statistical CSI, i.e., $f_{\bar{\mathbf{c}}}(\bar{c}_{\mathrm{s}},\bar{c}_{\mathrm{n}})$.  The following theorem reveals how the ML sequence can be obtained without instantaneous or statistical CSI if a full SCW code is employed.

\begin{theo}\label{Theo:ML_StrCW}
Assuming a \textit{full} SCW code is employed, i.e., $\mathbf{s}\in\boldsymbol{\mathcal{S}}^{\mathrm{sc}}(\bar{\boldsymbol{\omega}})$, the solutions of (\ref{Eq:ML}) and (\ref{Eq:ML_NonCoherent}) are identical and independent of both  \textit{instantaneous} CSI ($\bar{c}_{\mathrm{s}}$ and $\bar{c}_{\mathrm{n}}$) and  \textit{statistical} CSI ($f_{\bar{\mathbf{c}}}(\bar{c}_{\mathrm{s}},\bar{c}_{\mathrm{n}})$). This enables optimal CSI-free detection based on Algorithm~\ref{Alg:ML_StrCW}. Moreover, for a full \textit{binary} CW code, $\boldsymbol{\mathcal{S}}^{\mathrm{c}}(K,\omega)$,  the solution of (\ref{Eq:ML}) and (\ref{Eq:ML_NonCoherent}) is simply the codeword whose ``1" elements  correspond to the $\omega$ largest elements of $\mathbf{r}$. 
\end{theo}
\begin{IEEEproof}
The proof is provided in \iftoggle{ConfVersion}{the appendix}{Appendix~\ref{App:ML_StrCW}}.
\end{IEEEproof}

\begin{algorithm}[t] \label{Alg:ML_StrCW}
\caption{ML Sequence Detection for SCW Codes}
\begin{algorithmic}[1] 
\STATE \textbf{initialize} Sort observation vector $\mathbf{r}$ in ascending order into a new vector $\tilde{\mathbf{r}}$. 
\STATE Set those elements of $\mathbf{s}$ which correspond to the $\bar{\omega}_{0}$ first elements of $\tilde{\mathbf{r}}$ to $\eta_0=0$.
\FOR{$\ell=1$ until $\ell=L-1$}
        \STATE Set those elements of $\mathbf{s}$ which correspond to element $\sum_{\ell'=0}^{\ell-1}\bar{\omega}_{\ell'}+1$ to element $\sum_{\ell'=0}^{\ell-1}\bar{\omega}_{\ell'}+\bar{\omega}_{\ell}$ of $\tilde{\mathbf{r}}$ to $\eta_{\ell}$.
 \ENDFOR
 \STATE Return $\mathbf{s}$ as the ML sequence.
\end{algorithmic}
\end{algorithm}

We note that the ML sequence is not necessarily unique, i.e., more than one sequence may achieve the maximum value of the likelihood function in (\ref{Eq:ML}) and (\ref{Eq:ML_NonCoherent}). This can be also seen from Algorithm~\ref{Alg:ML_StrCW} where the ordered vector $\tilde{\mathbf{r}}$ may not necessarily be unique since some elements of $\mathbf{r}$ can be identical.   To further explain the optimal sequence detector for SCW codes in Algorithm~\ref{Alg:ML_StrCW}, we present the following examples.

\begin{examp} Suppose an SCW code with symbol set $\mathcal{S}=\{0,0.5,1\}$ and weight vector $\bar{\boldsymbol{\omega}}=[2, 3, 1]^{\mathsf{T}}$ is employed and we wish to decode the observation vector $\mathbf{r}=[12, 4, 8, 6, 15, 10]^{\mathsf{T}}$. 

\begin{itemize}
\item In line 1 of Algorithm~\ref{Alg:ML_StrCW}, $\mathbf{r}$ is reordered in ascending order into vector $\tilde{\mathbf{r}}=[4, 6, 8, 10, 12, 15]^{\mathsf{T}}$.
\item In line 2 of Algorithm~\ref{Alg:ML_StrCW}, the two elements ($\bar{\omega}_{0}=2$) of $\mathbf{s}$ corresponding to the first two elements of $\tilde{\mathbf{r}}$ are set to $\eta_0=0$.  This leads to $\mathbf{s}= [\times,0,\times,0,\times,\times]^{\mathsf{T}}$.
\item In line 4 of Algorithm~\ref{Alg:ML_StrCW}, the three elements ($\bar{\omega}_{1}=3$) of $\mathbf{s}$ corresponding to the third to the fifth elements of $\tilde{\mathbf{r}}$ are set to $\eta_1=0.5$.  This leads to $\mathbf{s}= [0.5,0,0.5,0,\times,0.5]^{\mathsf{T}}$.
\item In line 4 of Algorithm~\ref{Alg:ML_StrCW}, the one remaining element ($\bar{\omega}_{2}=1$) of $\mathbf{s}$ corresponding to the sixth element of $\tilde{\mathbf{r}}$ is set to $\eta_2=1$. This leads to the ML sequence $\mathbf{s}= [0.5,0,0.5,0,1,0.5]^{\mathsf{T}}$ which is returned in line 6 of Algorithm~\ref{Alg:ML_StrCW}.
\end{itemize}   
\end{examp} 

\begin{examp} Suppose a balanced binary CW code of length $K=6$, i.e., $\mathcal{S}=\{0,1\}$ and $\omega=3$, is employed and we wish to decode the observation vector $\mathbf{r}=[12, 4, 8, 6, 15, 8]^{\mathsf{T}}$. According to Theorem~\ref{Theo:ML_StrCW}, the optimal sequence is  the codeword whose ``1" elements  correspond to the $\omega=3$ largest elements of $\mathbf{r}$, i.e., elements $15$, $12$, and $8$. However, since we have two elements with value $8$, we obtain two ML sequences as $\mathbf{s}= [1,0,0,0,1,1]^{\mathsf{T}},[1,0,1,0,1,0]^{\mathsf{T}}$.  
\end{examp} 

\begin{remk}
  We note that for increasing codeword length, $K$, the length of observation vector $\mathbf{r}$, which needs to be sorted into $\tilde{\mathbf{r}}$, and the number of assignment operations in each iteration of the for-loop in Algorithm~\ref{Alg:ML_StrCW}, \textit{proportionally} increase. Therefore, the complexity of Algorithm~\ref{Alg:ML_StrCW} is linear in the codeword length, $K$. Based on the \textit{Van Emde Boas tree}, the sorting operation can be performed with a complexity on the order of $O(K\mathsf{log}(\mathsf{log}(L)))$ \cite{SortingComplexity}. Note that the complexity is exponential in $K$ for the general coherent and non-coherent ML problems in (\ref{Eq:ML}) and (\ref{Eq:ML_NonCoherent}), since the number of codewords and hence, the number of metrics which have to be computed, grow exponentially in $K$. Therefore, adopting the proposed SCW codes not only avoids the complexity and challenges of CSI acquisition but also significantly reduces the complexity of ML detection. This makes SCW codes particularly suitable for simple nano-machines with limited computational capabilities.
\end{remk}

While Theorem~\ref{Theo:ML_StrCW} claims CSI-free detection for full SCW codes,  in the following, we show that for binary CW codes, CSI-free detection is possible even if the codebook is not full.  

\begin{corol}\label{Corol:BinaryCW}
For \textit{binary} CW codes (not necessarily full codes), i.e., $\mathbf{s}\in\boldsymbol{\mathcal{S}}^{\mathrm{c}}(K,\omega)$ and $\mathcal{S}=\{0,1\}$, the solutions of (\ref{Eq:ML}) and (\ref{Eq:ML_NonCoherent}) are identical and require neither instantaneous CSI  nor statistical CSI. In this case, the optimal CSI-free decision is obtained from
\begin{align} \label{Eq:ML_BinSol}
  \hat{\mathbf{s}} &= \underset{\mathbf{s}\in\boldsymbol{\mathcal{S}}^{\mathrm{c}}(K,\omega)}{\mathrm{argmax}} \,\, \omega_{1}(\mathbf{s},\mathbf{r}) = \underset{\mathbf{s}\in\boldsymbol{\mathcal{S}}^{\mathrm{c}}(K,\omega)}{\mathrm{argmax}} \,\,\sum_{k=1}^K s[k]r[k].
\end{align}
\end{corol}
 \begin{IEEEproof}
The proof  follows directly from substituting binary symbols, i.e., $\mathcal{S}=\{0,1\}$, into (\ref{Eq:ML_Sol}) in \iftoggle{ConfVersion}{the Appendix}{Appendix~\ref{App:ML_StrCW}}.
 \end{IEEEproof}

\section{Performance Analysis}
In this section, we analyze the code rate and error rate of the proposed SCW codes. 

 \subsection{Rate Analysis}
 
The rate of a general code comprised of $M$ codewords of length $K$ with symbol set $\mathcal{S}$ is given by 
\begin{IEEEeqnarray}{lll} \label{Eq:CodeRateGen}
 R^{\mathrm{code}}(\bar{\boldsymbol{\omega}}) = \frac{\mathsf{log}\left(M\right)}{\mathsf{log}\left(|\mathcal{S}|^K\right)} = \frac{\mathsf{log}_{|\mathcal{S}|}\left(M\right)}{K}.
\end{IEEEeqnarray}
The code rate of a full SCW code is an upper bound for the code rate of SCW codes that do not use all possible codewords. Hence, in the following, we consider the code rate of the full SCW~codes.
 
 \begin{prop}\label{Prop:CodeRate}
 The code rate of a full SCW code, $\boldsymbol{\mathcal{S}}^{\mathrm{sc}}(\bar{\boldsymbol{\omega}})$, is given by
\begin{IEEEeqnarray}{lll} \label{Eq:StrCT_Rate}
R^{\mathrm{code}}(\bar{\boldsymbol{\omega}})&=\frac{1}{\sum_{\ell=1}^K \bar{\omega}_{\ell}}\sum_{\ell =0}^{L-1} \mathsf{log}_{L}\left({\sum_{\ell'\leq\ell} \bar{\omega}_{\ell'}\choose \bar{\omega}_{\ell}}\right) \nonumber \\
&=  \frac{1}{K}\mathsf{log}_{L}\left(\frac{K!}{\prod_{\ell=0}^{L-1}\bar{\omega}_{\ell}!}\right) \overset{K\to\infty}{\to}  H_L(\boldsymbol{\rho}),
\end{IEEEeqnarray}
where $\boldsymbol{\rho}=[\rho_0,\rho_1,\dots,\rho_L]^{\mathsf{T}}$ and $\rho_{\ell}=\bar{\omega}_{\ell}/K$.
 \end{prop}
 \begin{IEEEproof}
 \iftoggle{ConfVersion}{%
The proof is provided in \cite[Appendix~B]{ISIT17}.
}{%
The proof is provided in Appendix~\ref{App:PropRate}.
}
 \end{IEEEproof}

Given $K$ and $L$, the code rate of SCW codes is maximized when they are balanced, i.e., $\bar{\omega}_{\ell}=\bar{\omega}_{\ell'},\,\,\forall \ell,\ell'$ assuming $K/L$ is an integer. Moreover, for balanced codes, the rate approaches $R^{\mathrm{code}}(\bar{\boldsymbol{\omega}})\to 1$ as $K\to\infty$. We note that the code rate specifies the information content of a codeword compared to uncoded transmission with the same symbol set. Therefore, the code rate in (\ref{Eq:CodeRateGen}) is unitless. Alternatively, one can define the data rate in bits/symbol  as the average number of information bits that a symbol in a codeword contains.


\subsection{Error Analysis}
The average codeword error rate (CER) is denoted by $\bar{P}_e^{\mathrm{code}}(\bar{\boldsymbol{\omega}})=\mathsf{E}_{\bar{\mathbf{c}}}\{P_e^{\mathrm{code}}(\bar{\boldsymbol{\omega}}|\bar{\mathbf{c}})\}$  where $P_e^{\mathrm{code}}(\bar{\boldsymbol{\omega}}|\bar{\mathbf{c}})$ is the CER of the SCW code with weight $\bar{\boldsymbol{\omega}}$ for a given realization of the CSI $\bar{\mathbf{c}}$. In the following, we provide several analytical bounds for the CER $P_e^{\mathrm{code}}(\bar{\boldsymbol{\omega}}|\bar{\mathbf{c}})$.  First, we present an upper bound on the CER based on the pairwise error probability (PEP) and union and Chernoff bounds.

\begin{prop}\label{Prop:UpperGen}
The CER of the optimal detector for SCW codes, $\boldsymbol{\mathcal{S}}^{\mathrm{sc}}(\bar{\boldsymbol{\omega}})$,  is upper bounded~by
\begin{IEEEeqnarray}{lll} \label{Eq:UpperGen}
P_e^{\mathrm{code}}(\bar{\boldsymbol{\omega}}|\bar{\mathbf{c}})  \\
 \leq  \frac{1}{M} \sum_{\forall\mathbf{s}} \sum_{\forall\hat{\mathbf{s}}\neq \mathbf{s}} \mathsf{exp}\left(\sum_{k=1}^K \lambda[k]\left(\mathsf{exp}\left(\varpi[k]t\right)-1\right)\right),\,\,\forall t>0,\nonumber
\end{IEEEeqnarray}
where $\lambda[k]=s[k]\bar{c}_{\mathrm{s}}+\bar{c}_{\mathrm{n}}$ and $\varpi[k]=\ln\left( \frac{1+ \hat{s}[k] \mathsf{SNR}} {1+ s[k] \mathsf{SNR}} \right)$. In (\ref{Eq:UpperGen}), $t$ is an arbitrary positive real number which is introduced by the Chernoff bound that was used to arrive at (\ref{Eq:UpperGen}).
\end{prop}
 \begin{IEEEproof}
  \iftoggle{ConfVersion}{%
The proof is provided in \cite[Appendix~C]{ISIT17}.
}{%
The proof is provided in Appendix~\ref{App:PropUpperGen}.
}
 \end{IEEEproof}

 We note that (\ref{Eq:UpperGen}) constitutes an upper bound on the CER for any value of $t>0$. Therefore, one can optimize $t$ to tighten the upper bound. In the following corollary, we present a tighter upper bound than the general upper bound presented in Proposition~\ref{Prop:UpperGen} for binary~CW~codes. For notational simplicity,  we enumerate the codewords by $\mathbf{s}_i,\,\,i=1,\dots,M$. Moreover, let $d_{ij}=h(\mathbf{s}_i,\mathbf{s}_j)$ be the Hamming distance between codewords $\mathbf{s}_i$ and $\mathbf{s}_j$.

\begin{corol}\label{Corol:CER_UppBin}
The CER of the optimal detector for binary CW code, $\boldsymbol{\mathcal{S}}^{\mathrm{c}}(K,\omega)$,  is upper bounded~by
\begin{IEEEeqnarray}{lll} 
P_e^{\mathrm{code}}(K,\omega |\bar{\mathbf{c}}) \leq \frac{1}{M} \sum_{\forall d_{ij},\,i\neq j} 0.5f_X(0)+\sum_{x=1}^{\infty}
 f_X(x),\quad\,\, \label{Eq:CER_UppBin}
\end{IEEEeqnarray}
where $f_X(x)$ is given by 
\begin{IEEEeqnarray}{lll} 
 f_X(x) = e^{-(\lambda_1+\lambda_2)}\left(\frac{\lambda_2}{\lambda_1}\right)^{x/2}I_x(2\sqrt{\lambda_1\lambda_2}), \label{Eq:Skellam}
\end{IEEEeqnarray}
with $\lambda_1=\frac{d_{ij}(\bar{c}_{\mathrm{s}}+\bar{c}_{\mathrm{n}})}{2}$, $\lambda_2=\frac{d_{ij}\bar{c}_{\mathrm{n}}}{2}$, and $I_x(\cdot)$ is the modified Bessel function of the first kind and order $x$ \cite{TableIntegSerie}.
\end{corol}
 \begin{IEEEproof}
  \iftoggle{ConfVersion}{%
The proof is provided in \cite[Appendix~D]{ISIT17}.
}{%
The proof is provided in Appendix~\ref{App:Corol_CER_UppBin}.
}
\end{IEEEproof}

The upper bounds in Proposition~\ref{Prop:UpperGen} and Corollary~\ref{Corol:CER_UppBin} are based on the PEP and the union bound. Hence, they are expected to be tight at high SNRs. In the following proposition, we provide upper and lower bounds on the CER for the special case of \textit{full} binary CW codes which are tight for all SNRs.

\begin{prop}\label{Prop:UppBinFull}
The CER of the optimal detector for a full binary CW code, $\boldsymbol{\mathcal{S}}^{\mathrm{c}}(K,\omega)$,~is~bounded~as
\begin{IEEEeqnarray}{lll} 
\Scale[1]{\displaystyle \sum_{y=1}^{\infty} \hspace{-0.5mm} F_X(y-1)f_Y(y) 
\leq P_e^{\mathrm{code}}(K,\omega |\bar{\mathbf{c}})  
\leq \sum_{y=0}^{\infty} \hspace{-0.5mm} F_X(y)f_Y(y),\quad\,\,} \,\, \label{Eq:CER_sum}
\end{IEEEeqnarray}
where $F_X(\cdot)$ and $f_Y(\cdot)$ are given by
\begin{IEEEeqnarray}{rll} \label{Eq:PDFMaxMin}
F_X(x) &\,\,= 1- (1-F_{\mathcal{P}}(x,\bar{c}_{\mathrm{s}}+\bar{c}_{\mathrm{n}}))^{\omega} \qquad \IEEEyesnumber \IEEEyessubnumber \\
f_Y(y) &\,\,= (K-\omega) f_\mathcal{P}(y,\bar{c}_{\mathrm{n}}) F_{\mathcal{P}}(y,\bar{c}_{\mathrm{n}})^{K-\omega -1}.  \IEEEyessubnumber 
\end{IEEEeqnarray}
In (\ref{Eq:PDFMaxMin}a) and (\ref{Eq:PDFMaxMin}b), $f_\mathcal{P}(\cdot,\cdot)$ and $F_{\mathcal{P}}(\cdot,\cdot)$ are given by
\begin{IEEEeqnarray}{rll} \label{Eq:PDFPoisson}
f_\mathcal{P}(x,\lambda) &= \frac{\lambda^{x}e^{-\lambda}}{x!} \qquad \IEEEyesnumber \IEEEyessubnumber \\
F_\mathcal{P}(x,\lambda) &= Q(\lfloor x+1\rfloor,\lambda),  \IEEEyessubnumber 
\end{IEEEeqnarray}
where $Q(\cdot,\cdot)$ is the regularized Gamma function \cite{TableIntegSerie}.
\end{prop}
 \begin{IEEEproof}
   \iftoggle{ConfVersion}{%
The proof is provided in \cite[Appendix~E]{ISIT17}.
}{%
The proof is provided in Appendix~\ref{App:PropUppBinFull}.
}
\end{IEEEproof}

\section{Performance Evaluation}

\begin{figure*}[!tbp]
  \centering
  \begin{minipage}[b]{0.31\textwidth}
  \centering

\resizebox{1.1\linewidth}{!}{\hspace{-0.7cm}\psfragfig{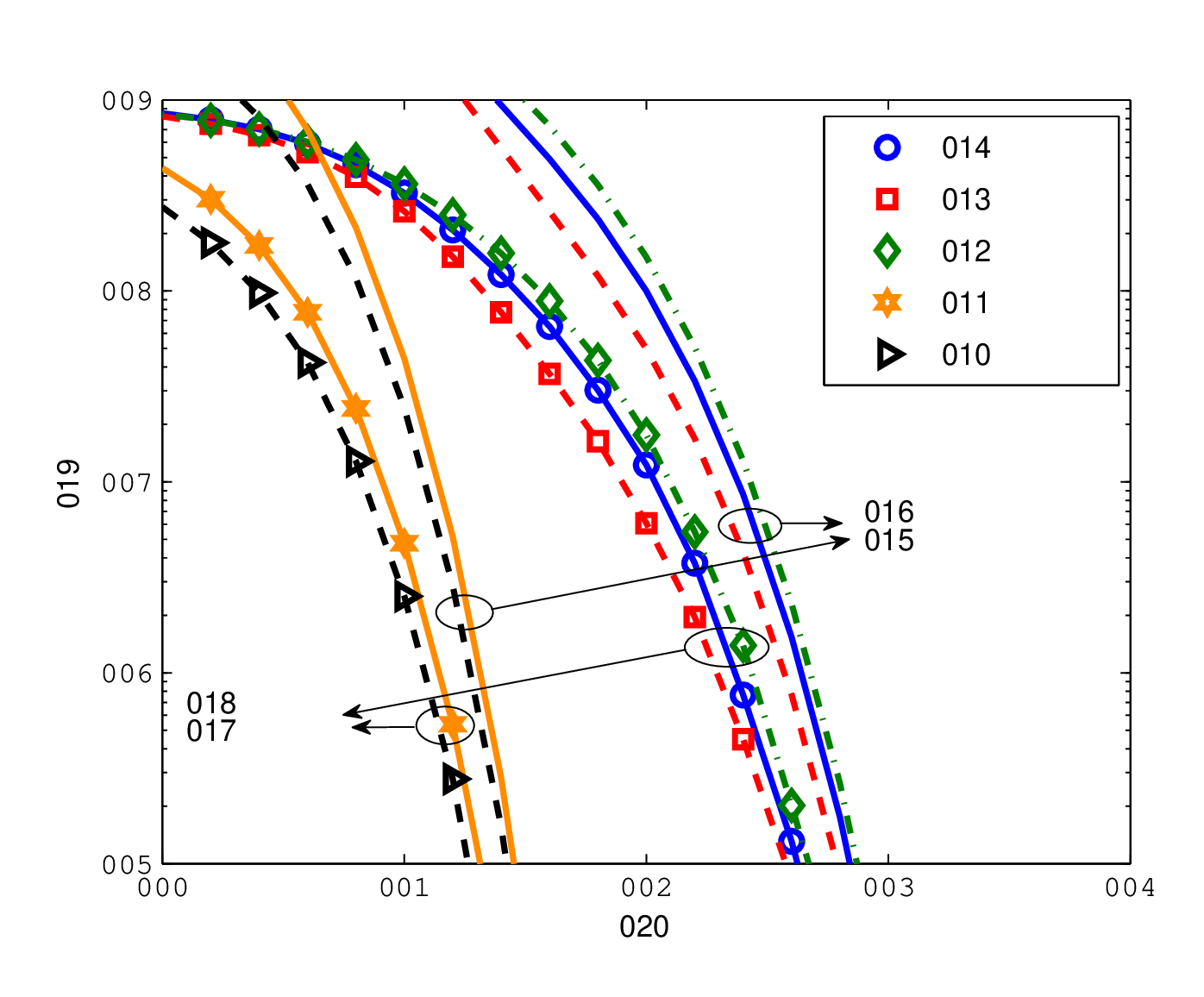}} \vspace{-0.8cm}
\caption{CER for SCW codes, $P_e^{\mathrm{code}}(\bar{\boldsymbol{\omega}})$, versus the SNR in dB for different weights $\bar{\boldsymbol{\omega}}$. \vspace{-0.01cm} }
\label{Fig:CER_SCW}
  \end{minipage}
    \hfill
  \begin{minipage}[b]{0.01\textwidth}
  \end{minipage}
  \hfill
  \begin{minipage}[b]{0.31\textwidth}
  \centering
\resizebox{1.1\linewidth}{!}{\hspace{-0.6cm}\psfragfig{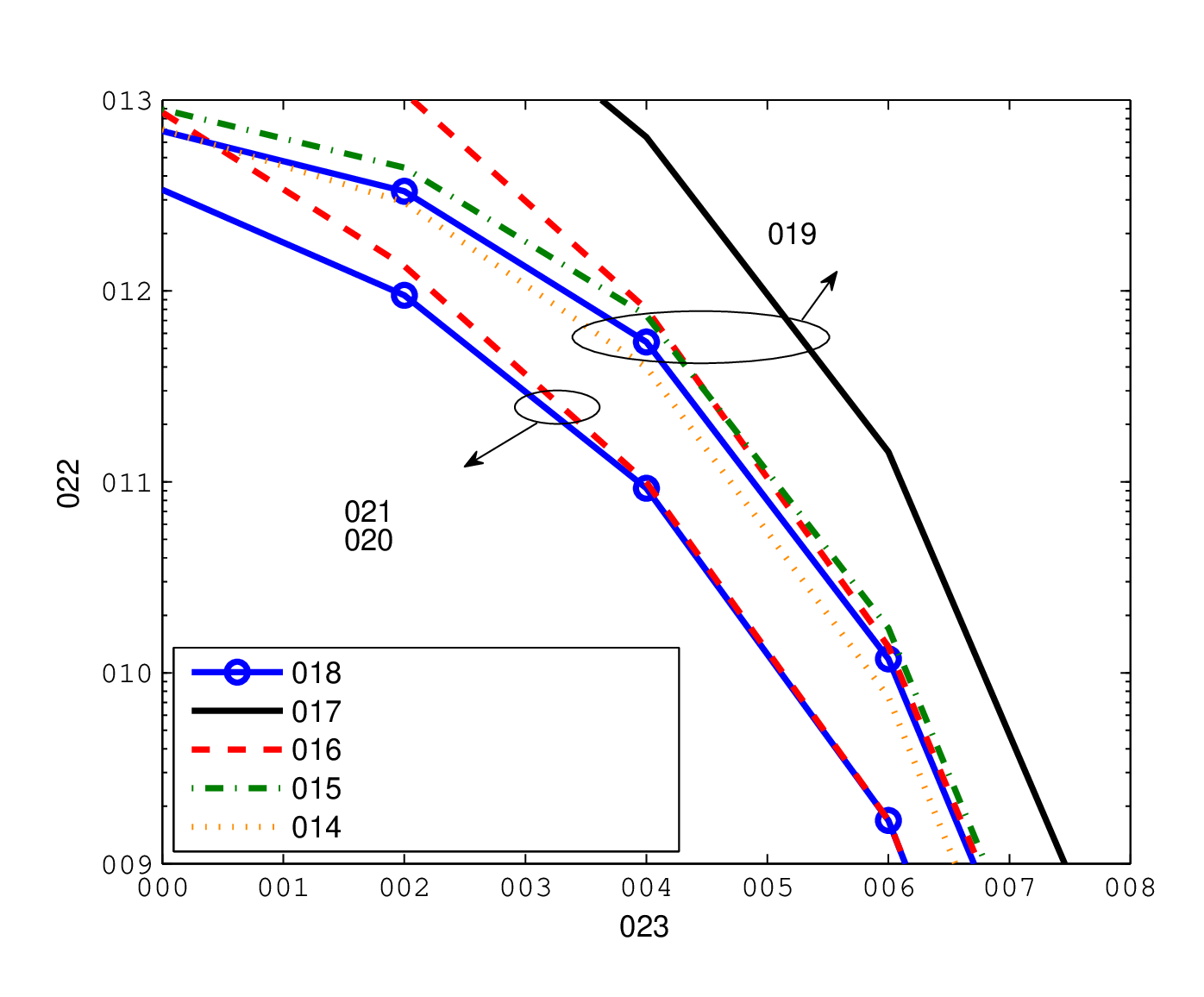}} \vspace{-0.8cm}
\caption{CER for binary CW codes, $P_e^{\mathrm{code}}(K,\omega)$, versus the SNR in dB for $K=10$ and $\rho=\frac{1}{2}$.\vspace{-0.2cm} }
\label{Fig:CER_SNR}
  \end{minipage}
    \hfill
  \begin{minipage}[b]{0.01\textwidth}
  \end{minipage}\vspace{-0.2cm}
  \hfill
  \begin{minipage}[b]{0.31\textwidth}
  \centering
\resizebox{1.1\linewidth}{!}{\hspace{-0.7cm}\psfragfig{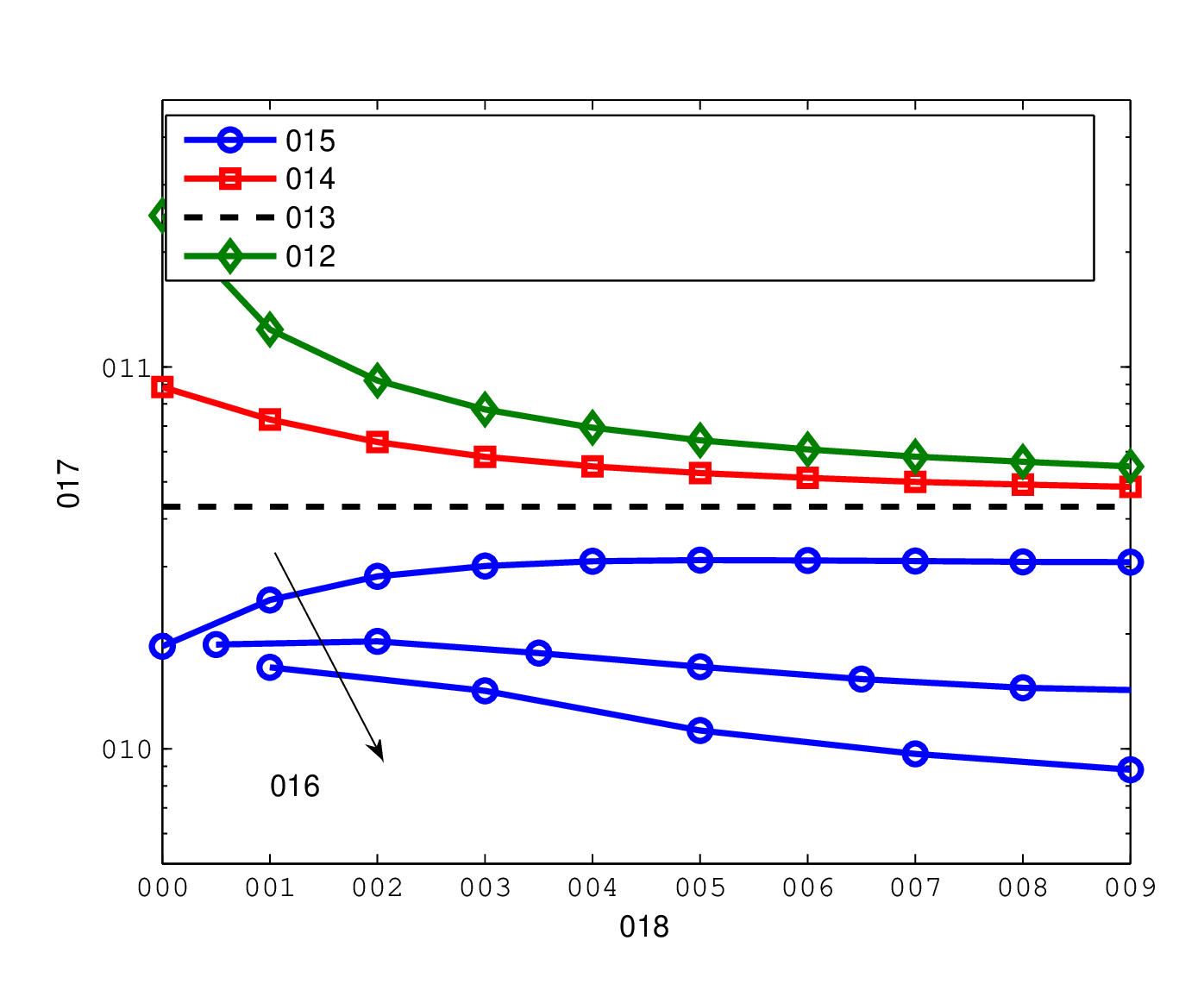}} \vspace{-0.8cm}
\caption{BER versus the codeword/block length, $K$, for $\rho=\frac{1}{2}$, $\mathrm{SNR}=5$ dB, and $R\in\{\frac{1}{2},\frac{1}{3},\frac{1}{4}\}$.\vspace{-0.2cm} }
\label{Fig:BER_SNR}
  \end{minipage}
    \hfill
  \begin{minipage}[b]{0.01\textwidth}
  \end{minipage}\vspace{-0.2cm}
\end{figure*}

Since the proposed detection scheme does not require CSI, it can be adopted regardless of the
channel being deterministic/time-invariant or stochastic/time-variant. In Figs.~\ref{Fig:CER_SCW} and \ref{Fig:CER_SNR}, we adopt the deterministic channel with flow introduced in \cite{Adam_Enzyme},  and in Fig.~\ref{Fig:BER_SNR}, we use the stochastic channel in \cite{NanoCOM16}. Due to space constraints, we avoid restating the details of these channel models and refer the readers to \cite{NanoCOM16,Adam_Enzyme} for detailed descriptions. In particular, the models in \cite{NanoCOM16,Adam_Enzyme} are based on the following equation for the \textit{expected} number of molecules observed at the receiver as a function of time 
\begin{align} \label{Eq:Consentration}
\hspace{-0.25cm}  \bar{c}_{\mathrm{s}}(t) = \frac{N^{\mathrm{tx}}V^{\mathrm{rx}}}{(4\pi D t)^{3/2}} \mathrm{exp}\left(-\kappa\bar{c}_{\mathrm{e}}t-\frac{(d-v_{\parallel}t)^2+(v_{\perp}t)^2}{4Dt}\right), \hspace{-0.15cm}
\end{align}
where the definition of the involved variables and their default values are provided in Table~I. We assume a sampling time of $T^{\mathrm{samp}}=0.1$ ms and a symbol duration of $T^{\mathrm{symb}}=1$ ms. For instance, for the default values of the system parameters given in Table~I, we obtain $\bar{c}_{\mathrm{s}}= \bar{c}_{\mathrm{s}}(T^{\mathrm{samp}})=4.9$ molecules. Moreover, the expected number of noise molecules can be determined based on the adopted SNR value according to $\bar{c}_{\mathrm{n}}= \bar{c}_{\mathrm{s}} \mathsf{SNR}^{-1}$. Alternatively, for a fixed $\bar{c}_{\mathrm{n}}$, one may change the number of released molecules, $N^{\mathrm{tx}}$, to obtain different SNRs. Here, we adopt the latter approach with $\bar{c}_{\mathrm{n}}=4.9$, i.e., the SNR using the default values of the system parameters in Table~I is zero dB. Finally, for the simulation results provided in this section, we choose symbol set  $\mathcal{S}=\left\{0,\frac{1}{L-1},\frac{2}{L-1},\cdots,\frac{L-2}{L-1},1\right\}$.

\begin{table}
\label{Table:Parameter}
\caption{Default Values of the System Parameters \cite{NanoCOM16,Adam_Enzyme}. \vspace{-0.2cm}} 
\begin{center}
\scalebox{0.59}
{
\begin{tabular}{|| c | c | c ||}
  \hline 
  Variable & Definition & Value \\ \hline \hline
       $N^{\mathrm{tx}}$ & Number of released molecules  & $10^4$ molecules \\ \hline      
       $V^{\mathrm{rx}}$ & Receiver volume   & $\frac{4}{3}\pi 50^3$ \,\, ${\text{nm}}^3$ \\  
         &    & (a sphere with radius $50$ nm) \\ \hline   
        $d$ &  Distance between the transmitter and the receiver  & $500$ nm\\ \hline 
         $D$ &  Diffusion coefficient for the signaling molecule & $4.3\times 10^{-10}$ $\text{m}^2\cdot\text{s}^{-1}$\\ \hline          
       $\bar{c}_{\mathrm{e}}$ &  Enzyme concentration  & $10^{5}$ $\text{molecule}\cdot\mu\text{m}^3$ \\   
         &     & (approx. $1.66$ micromolar) \\ \hline  
       $\kappa$ &    Rate of molecule degradation reaction & $2\times10^{-19}$ $\text{m}^3\cdot\text{molecule}^{-1}\cdot\text{s}^{-1}$ \\ \hline 
       $(v_{\parallel},v_{\perp})$ &  Components of flow velocity   & $(10^{-3},10^{-3})$ $\text{m}\cdot\text{s}^{-1}$ \\ \hline
\end{tabular}
}
\end{center}\vspace{-0.5cm}
\end{table}

Next, we evaluate the error performance of the proposed CSI-free detector. To examine the performance of different SCW codes, we adopt a simple ternary symbol set, i.e., $\mathcal{S}=\{0,0.5,1\}$, and a small codeword length, i.e., $K=6$. Moreover, we consider the following five weights: $\bar{\boldsymbol{\omega}}=[2,2,2]^{\mathsf{T}}$ which yields a balanced code, $\bar{\boldsymbol{\omega}}=[3,2,1]^{\mathsf{T}},[1,2,3]^{\mathsf{T}}$ which yield unbalanced codes, $\bar{\boldsymbol{\omega}}=[3,0,3]^{\mathsf{T}}$ which is equivalent to a binary balanced code, and $\bar{\boldsymbol{\omega}}=[5,0,1]^{\mathsf{T}}$ which is equivalent to pulse position modulation (PPM) \cite{PPM_Pieroborn}. In Fig.~\ref{Fig:CER_SCW}, we show the CER for these SCW codes, $P_e^{\mathrm{code}}(\bar{\boldsymbol{\omega}})$, versus the SNR in dB. In addition, we plot the upper bound given in Proposition~\ref{Prop:UpperGen} for $t=0.5$ \footnote{For simplicity, we choose a fixed $t$ for the results shown in Figs.~\ref{Fig:CER_SCW} and \ref{Fig:CER_SNR}, i.e., $t=0.5$. This specific value of $t$ was chosen in a trial-and-error manner  without claim of optimality in terms of the tightness of the upper bound.}. Fig.~\ref{Fig:CER_SCW} confirms the validity of the proposed upper bound and that it becomes tighter at high SNRs. We note that comparing the curves in  Fig.~\ref{Fig:CER_SCW} is not entirely fair. The common properties of the  codes considered in this figure include that CSI is not needed for detection and they employ the same codeword length, $K$, an identical  per-symbol ``power" constraint, $N^{\mathrm{tx}}$, and \textit{in principle} the same symbol set, $\mathcal{S}$. However, their code rates, $R^{\mathrm{code}}(\bar{\boldsymbol{\omega}})$, and average power consumption, $\bar{N}^{\mathrm{tx}}$, are not necessarily identical. For instance, the binary balanced codes achieve a lower CER compared to the ternary balanced code at the cost of a lower achievable~data~rate.

 The SCW codes adopted in Fig.~\ref{Fig:CER_SCW} are full codes, i.e., all possible codewords are used. In Corollary~\ref{Corol:BinaryCW}, we showed that CSI-free detection is possible also for binary CW codes with partial codebooks. In Fig.~\ref{Fig:CER_SNR}, we show the CER for binary CW codes, $P_e^{\mathrm{code}}(K,\omega)$, versus the SNR in dB for $K=10$ and $\rho=\frac{1}{2}$. The results for both the partial code with code rate $R=0.5$ and the full code with rate $R(K,\omega)=0.8$ are included. In particular, to generate the partial codebook, $2^{0.5K}=32$ codewords are randomly chosen out of all $M=252$ possible codewords. We observe that the code with partial codebook achieves a lower CER at the expense of a lower code rate. In addition, in Fig.~\ref{Fig:CER_SNR}, we show the upper bounds proposed in Proposition~\ref{Prop:UpperGen},  Corollary~\ref{Corol:CER_UppBin}, and Proposition~\ref{Prop:UppBinFull} and the  lower bound  proposed in Proposition~\ref{Prop:UppBinFull}. We note that  the bounds in Proposition~\ref{Prop:UppBinFull} are valid only for full codes. Fig.~\ref{Fig:CER_SNR} confirms the validity of the bounds and that the upper bounds proposed in  Proposition~\ref{Prop:UppBinFull} and  Corollary~\ref{Corol:CER_UppBin} for the binary CW codes are tighter than the upper bound proposed in Proposition~\ref{Prop:UpperGen} for general SCW codes. Moreover, Fig.~\ref{Fig:CER_SNR} reveals that the bounds in Proposition~\ref{Prop:UppBinFull} are fairly tight for all SNRs whereas the upper bound in Corollary~\ref{Corol:CER_UppBin} is particularly tight at high SNRs.

 Finally, in Fig.~\ref{Fig:BER_SNR}, we compare the proposed coded communication scheme with an uncoded transmission employing the coherent symbol-by-symbol detector in \cite{HamidJSAC} and the optimal non-coherent and the sub-optimal CSI-free detectors in \cite{NanoCOM16}. In Fig.~\ref{Fig:BER_SNR},  we show the bit error rate (BER) versus the codeword  length (or the ``block" length in \cite{NanoCOM16}), $K$, for $\rho=\frac{1}{2}$, $\mathrm{SNR}=5$ dB, $R\in\{\frac{1}{2},\frac{1}{3},\frac{1}{4}\}$ and Scenario~2 of the stochastic MC channel in \cite{NanoCOM16}. As previously confirmed in \cite{NanoCOM16}, the BERs of the optimal non-coherent and the sub-optimal CSI-free detectors approach that of the optimal coherent detector as $K\to\infty$. The proposed CSI-free detector based on SCW codes outperforms all considered uncoded benchmark schemes at the expense of a lower rate. Furthermore, the performance gain of the proposed coded communication over the uncoded benchmark schemes increases as the code rate decreases.
 
\section{Conclusion}

We have proposed SCW codes which facilitate \textit{optimal CSI-free} sequence detection at the cost of decreasing the data rate compared to  uncoded transmission. For the proposed SCW codes, we analyzed the code rate and the error rate. Simulation results verified our analytical derivations and showed that the proposed CSI-free detector for SCW codes outperforms the optimal coherent and non-coherent detectors for uncoded transmission.

\iftoggle{ConfVersion}{%

\appendix
  
}{%

\appendices  
\section{Proof of Theorem~\ref{Theo:ML_StrCW}}
  \label{App:ML_StrCW}
  
}

First, we rewrite (\ref{Eq:ML}) as follows
\begin{align} \label{Eq:ML_Sol} 
  \hat{\mathbf{s}}  & \overset{(a)}{=} \underset{\mathbf{s}\in\boldsymbol{\mathcal{S}}}{\mathrm{argmax}} \,\,\mathsf{ln}\left(f_{\mathbf{r}}(\mathbf{r}|\bar{\mathbf{c}},\mathbf{s})\right)  \nonumber  \\
  & = \underset{\mathbf{s}\in\boldsymbol{\mathcal{S}}}{\mathrm{argmax}} \, \sum_{k=1}^{K} r[k] \mathsf{ln}\left( \bar{c}_{\mathrm{s}}s[k]+\bar{c}_{\mathrm{n}}\right) - \bar{c}_{\mathrm{s}}s[k]-\bar{c}_{\mathrm{n}} - \mathsf{ln}\left(r[k]!\right)
   \nonumber \\
  &  \overset{(b)}{=} \underset{\mathbf{s}\in\boldsymbol{\mathcal{S}}}{\mathrm{argmax}} \,\, 
    - \omega(\mathbf{s}) \bar{c}_{\mathrm{s}}  + \sum_{k=1}^{K} r[k] \mathsf{ln}\left( 1+ s[k] \mathsf{SNR}  \right),
\end{align}
where to arrive at equality $(a)$, we use the property that $\mathsf{ln}(\cdot)$ is a monotonically increasing function, and for equality $(b)$, we remove those terms that do not depend on the hypothesis sequence $\mathbf{s}$ and use the definitions of $\omega(\mathbf{s})$ and $\mathsf{SNR}$. 

For SCW codes, $\omega(\mathbf{s})$ is identical for all codewords and hence does not change the ML sequence. The second term in (\ref{Eq:ML_Sol}) is in fact  
a weighted sum of the observations $r[k]$ where the weights $\ln\left( 1+ s[k] \mathsf{SNR}  \right)$ are monotonically increasing functions of $s[k]$. Therefore, for the ML sequence $\mathbf{s}^*=[s^*[1],\dots,s^*[k]]^{\mathsf{T}}$, if $r[k]\geq r[k']$  holds, then $s^*[k]\geq s^*[k']$ has to hold. This leads to Algorithm~1 for general SCW codes.  For the case of binary CW codes, $\boldsymbol{\mathcal{S}}^{\mathrm{c}}(K,\omega)$, this leads to a sequence whose ``1" elements  correspond to the $\omega$ largest elements of $\mathbf{r}$.  The resulting sequence is optimal if it belongs to the codebook $\boldsymbol{\mathcal{S}}^{\mathrm{sc}}(\bar{\boldsymbol{\omega}})$. This condition is ensured if the code is full. 
Note that this is the solution of the ML problem in (\ref{Eq:ML}) for coherent sequence detection. If for a given CSI $(\bar{c}_{\mathrm{s}},\bar{c}_{\mathrm{n}})$, the sequence $\mathbf{s}^*$  that maximizes the conditional likelihood function $f_{\mathbf{r}}(\mathbf{r}|\bar{\mathbf{c}},\mathbf{s})$ does not depend on the CSI value, the average likelihood function in (\ref{Eq:ML_NonCoherent}) is also maximized by $\mathbf{s}^*$. In other words, the solutions of (\ref{Eq:ML}) and (\ref{Eq:ML_NonCoherent}) for coherent and non-coherent detection are identical and do not depend on instantaneous nor statistical CSI. Therefore, an SCW code enables optimal CSI-free detection.  These results are concisely summarized in Theorem~\ref{Theo:ML_StrCW} and Algorithm~1 which concludes~the~proof.

\iftoggle{ConfVersion}{%

}{%

\section{Proof of Proposition~\ref{Prop:CodeRate}}\label{App:PropRate}

In the following, using (\ref{Eq:CodeRateGen}), we derive the code rate of a \textit{full} SCW code. First, note that we have $|\mathcal{S}|=L$ and $K=\sum_{\ell=0}^{L-1}\bar{\omega}_{\ell}$ for SCW code $\boldsymbol{\mathcal{S}}^{\mathrm{sc}}(\bar{\boldsymbol{\omega}})$. In order to find $M$ for a given SCW code $\boldsymbol{\mathcal{S}}^{\mathrm{sc}}(\bar{\boldsymbol{\omega}})$, we use the definition of the binomial coefficient, i.e., ${n\choose k}=\frac{n!}{k!(n-k)!}$. In particular, there are ${K\choose \bar{\omega}_{L-1}}$ possibilities for the positions of symbol $\eta_{L-1}=1$. Having the positions of symbol $\eta_{L-1}$ fixed, there are ${K-\bar{\omega}_{L-1}\choose \bar{\omega}_{L-2}}$ possibilities for the positions of symbol $\eta_{L-2}$. Continuing this process, we obtain $M$ for a full SCW code $\boldsymbol{\mathcal{S}}^{\mathrm{sc}}(\bar{\boldsymbol{\omega}})$ as
\begin{IEEEeqnarray}{lll} \label{Eq:StrCT_M}
M&={K\choose \bar{\omega}_{L-1}}{K-\bar{\omega}_{L-1}\choose \bar{\omega}_{L-2}}
\cdots{\bar{\omega}_{0}+\bar{\omega}_{1}\choose \bar{\omega}_{1}} {\bar{\omega}_{0}\choose \bar{\omega}_{0}} \nonumber \\
&=\prod_{\ell =0}^{L-1} {\sum_{\ell'\leq\ell} \bar{\omega}_{\ell'}\choose \bar{\omega}_{\ell}} 
= \frac{K!}{\prod_{\ell=0}^{L-1}\bar{\omega}_{\ell}!}.
\end{IEEEeqnarray}
Substituting the above results into (\ref{Eq:CodeRateGen}) leads to the first expression in (\ref{Eq:StrCT_Rate}). We note that the first expression in (\ref{Eq:StrCT_M}) is the well-known multinomial coefficient which can be written equivalently as the second expression in (\ref{Eq:StrCT_M}) \cite{TableIntegSerie}. Finally, we note that the entropy of an RV with multinomial distribution and probability vector $\boldsymbol{\rho}=[\rho_0,\rho_1,\dots,\rho_L]^{\mathsf{T}}$ where $\rho_{\ell}=\bar{\omega}_{\ell}/K$, asymptotically approaches $H_L(\boldsymbol{\rho})$ when $K\to\infty$ \cite{TableIntegSerie}. Therefore, we obtain $\mathsf{log}_L(M)\to KH_L(\boldsymbol{\rho})$ as $K\to\infty$. This leads to the asymptotic result in  (\ref{Eq:StrCT_Rate})  and concludes the proof.

\section{Proof of Proposition~\ref{Prop:UpperGen}}\label{App:PropUpperGen}

The PEP, denoted by $P(\mathbf{s}\to\hat{\mathbf{s}})$, is defined as the probability that assuming $\mathbf{s}$ is transmitted, $\hat{\mathbf{s}}$ is detected. Using the PEP, the CER is upper bounded based on the union bound as follows
\begin{IEEEeqnarray}{lll} \label{Eq:UB}
P_e^{\mathrm{code}}(\bar{\boldsymbol{\omega}}|\bar{\mathbf{c}}) & \leq \sum_{\forall\mathbf{s}} \sum_{\forall\hat{\mathbf{s}}\neq \mathbf{s}} P(\mathbf{s}\to\hat{\mathbf{s}}) \mathsf{Pr}(\mathbf{s}) \nonumber \\
 &\overset{(a)}{\leq}  \frac{1}{M} \sum_{\forall\mathbf{s}} \sum_{\forall\hat{\mathbf{s}}\neq \mathbf{s}} \mathsf{Pr}\{ X\geq 0\} \nonumber \\
 & \overset{(b)}{\leq} \frac{1}{M} \sum_{\forall\mathbf{s}} \sum_{\forall\hat{\mathbf{s}}\neq \mathbf{s}} 
  \mathsf{E}\left\{\mathsf{exp}\left(Xt\right)\right\} \nonumber \\
 & =\frac{1}{M} \sum_{\forall\mathbf{s}} \sum_{\forall\hat{\mathbf{s}}\neq \mathbf{s}}  G_X(t),\,\,\forall t>0,
\end{IEEEeqnarray}
where in inequality $(a)$, we use the property that the codewords are equiprobable, i.e., $\mathsf{Pr}(\mathbf{s})=\frac{1}{M}$, define $X=\Lambda^{\mathrm{ML}}(\hat{\mathbf{s}}) - \Lambda^{\mathrm{ML}}(\mathbf{s})$, and treat the case $X=0$ always as an error which upper bounds the PEP term $P(\mathbf{s}\to\hat{\mathbf{s}})$. For inequality $(b)$, we employ the Chernoff bound where $G_X(t)$ denotes the moment generating function (MGF) of RV $X$ \cite{StochGallager}. Using (\ref{Eq:ML_Sol}), $X$ can be rewritten as 
\begin{IEEEeqnarray}{lll} \label{Eq:X_weighted}
X = \sum_{k=1}^{K} r[k] \ln\left( \frac{1+ \hat{s}[k] \mathsf{SNR}}{1+ s[k] \mathsf{SNR}}  \right)
\triangleq \sum_{k=1}^{K} \varpi[k]r[k],
\end{IEEEeqnarray}
which is basically a weighted sum of the observations. Note that given $\mathbf{s}$, $r[k],\,\,\forall k$, is a Poisson RV with mean $\lambda[k]=s[k]\bar{c}_{\mathrm{s}}+\bar{c}_{\mathrm{n}}$ and MFG $G_{r[k]}(t)=\mathsf{exp}(\lambda[k](e^{t}-1))$. Exploiting the properties of MGFs, namely $G_{aX}(t)=G_X(at)$, where $a$ is a constant, and $G_{X+Y}(t)=G_{X}(t)G_Y(t)$ where $X$ and $Y$ are independent RVs, we obtain
\begin{IEEEeqnarray}{lll} \label{Eq:MGF}
G_X(t)=\prod_{k=1}^{K} G_{r[k]}\left(\varpi[k]t\right) = \mathsf{exp}\left(\sum_{k=1}^K \lambda[k]\left(e^{\varpi[k]t}-1\right)\right). \quad\,\,\,
\end{IEEEeqnarray}
The above result leads to the upper bound in (\ref{Eq:UpperGen}) and concludes the proof.

\section{Proof of Corollary~\ref{Corol:CER_UppBin}}\label{App:Corol_CER_UppBin}

Using the PEP, the CER is upper bounded based on the union bound as follows
\begin{IEEEeqnarray}{lll} \label{Eq:UB_Bin}
P_e^{\mathrm{code}}(\bar{\boldsymbol{\omega}}|\bar{\mathbf{c}}) & \leq \sum_{\forall\mathbf{s}} \sum_{\forall\hat{\mathbf{s}}\neq \mathbf{s}} P(\mathbf{s}\to\hat{\mathbf{s}}) \mathsf{Pr}(\mathbf{s}) \nonumber \\
 & = \frac{1}{M} \sum_{\forall\mathbf{s}} \sum_{\forall\hat{\mathbf{s}}\neq \mathbf{s}}  \mathsf{Pr}\{X>0\}+0.5\mathsf{Pr}\{X=0\}, \quad
\end{IEEEeqnarray}
where $X= \Lambda^{\mathrm{ML}}(\hat{\mathbf{s}}) - \Lambda^{\mathrm{ML}}(\mathbf{s})$. RV $X$ can be simplified as
\begin{IEEEeqnarray}{lll} 
X = \sum_{k=1}^K (\hat{s}[k]-s[k]) r[k] = \overset{X_2}{\overbrace{\sum_{k\in\widehat{\mathcal{K}}} \hat{s}[k]r[k]}}
 - \overset{X_1}{\overbrace{\sum_{k\in\mathcal{K}} s[k]r[k]}}, \quad
\end{IEEEeqnarray}
where $\mathcal{K}=\{k|s[k]=1\,\,\text{and}\,\,s[k]\neq\hat{s}[k]\}$ and $\widehat{\mathcal{K}}=\{k|\hat{s}[k]=1\,\,\text{and}\,\,s[k]\neq\hat{s}[k]\}$. Here, $X_1$ and $X_2$ are two \textit{independent} Poisson RVs with means $\lambda_1=\frac{d_{ij}(\bar{c}_{\mathrm{s}}+\bar{c}_{\mathrm{n}})}{2}$ and $\lambda_2=\frac{d_{ij}\bar{c}_{\mathrm{n}}}{2}$, respectively. Therefore, $X$ follows a Skellam distribution whose PDF is given in (\ref{Eq:Skellam}) \cite{Skellam}. Moreover,  since, for a given $\mathbf{s}$ and $\hat{\mathbf{s}}$, the Skellam distribution is a function of the Hamming distance $d_{ij}$, we can replace the summations in (\ref{Eq:UB_Bin}) by the summation over all $d_{ij}$ as in (\ref{Eq:CER_UppBin}). This completes the proof.

\section{Proof of Proposition~\ref{Prop:UppBinFull}}\label{App:PropUppBinFull} 
Let $\hat{\mathbf{s}}$ denote the detected codeword using the optimal detector. We divide the received vector $\mathbf{r}$ into two vectors $\tilde{\mathbf{r}}=[\tilde{r}_1,\tilde{r}_2,\dots,\tilde{r}_{\omega}]^T$ and $\hat{\mathbf{r}}=[\hat{r}_1,\hat{r}_2,\dots,\hat{r}_{K-\omega}]^T$ which correspond to positions of the ones and zeros in the transmitted codeword $\mathbf{s}$, respectively. Hereby, conditioned on $\mathbf{s}$, elements $\tilde{r}_i$ and $\hat{r}_j$ are independent Poisson RVs with means $\bar{c}_{\mathrm{s}}+\bar{c}_{\mathrm{n}}$ and $\bar{c}_{\mathrm{n}}$, respectively. Let us define $X=\mathsf{min}\{\tilde{r}_1,\tilde{r}_2,\dots,\tilde{r}_{\omega}\}$ and $Y=\mathsf{max}\{\hat{r}_1,\hat{r}_2,\dots,\hat{r}_{K-\omega}\}$. For the optimal detector and a full binary CW code, the CER is bounded as
\begin{IEEEeqnarray}{rll} \label{Eq:PoissMinMax}
\mathsf{Pr}\{X < Y \} \leq P_e^{\mathrm{code}}(K,\omega |\bar{\mathbf{c}})  \leq \mathsf{Pr}\{X \leq Y \}.
\end{IEEEeqnarray}
In fact, for events when $X=Y$ occurs, the detector selects with equal probability one of the hypothesis with the same value of $\Lambda^{\mathrm{ML}}(\mathbf{s})$. For the upper bound, we treat event $X=Y$ as an error and for the lower bound, we treat it as a correct decision. Using order statistics theory \cite{Order_Statistics}, the cumulative density function (CDF) of $X$ and the PDF of $Y$ are given by (\ref{Eq:PDFMaxMin}a) and (\ref{Eq:PDFMaxMin}b), respectively,  where $f_\mathcal{P}(\cdot,\lambda)$ and $F_{\mathcal{P}}(\cdot,\lambda)$ are in fact the PDF and CDF of a Poisson RV with mean $\lambda$, respectively \cite{Order_Statistics}. Using $F_X(x)$ and $f_Y(y)$, the lower and upper bounds in (\ref{Eq:PoissMinMax}) are rewritten in (\ref{Eq:CER_sum}). This completes the proof.

}

\section*{Acknowledgment}
This work was supported in part by the German Science Foundations (Project SCHO 831/7-1) and the Friedrich-Alexander-University Erlangen-N\"urnberg under the Emerging Fields Initiative (EFI).  The authors would like to thank Prof. Andrea Goldsmith for her valuable suggestions and comments for this~paper.

\bibliographystyle{IEEEtran}
\bibliography{Ref_27_04_2017}

\end{document}